\documentclass[acus]{JAC2000}

\usepackage{graphicx}

\begin{document}
\title{STUDIES OF BEAM OPTICS AND SCATTERING IN THE NEXT LINEAR
       COLLIDER POST-LINAC COLLIMATION SYSTEM
       \thanks{Work supported by
       U.S. Department of Energy, Contract DE-AC03-76SF00515}}

\author{P. Tenenbaum, R. Helm, L. Keller, T.O. Raubenheimer, 
        SLAC, Stanford, CA, USA}

\maketitle

\begin{abstract} 
We present a new conceptual and optical design for the Next Linear
Collider post-linac collimation system.  Energy collimation and
passive protection against off-energy beams are achieved in a
system with large horizontal dispersion and vertical 
betatron functions. 
Betatron collimation is performed in a relatively low-beta
(FODO-like) lattice in which only thin spoilers intercept 
particles near the beam core, while thick absorbers maintain
a large stay-clear from the beam.  Two possible schemes for the
spoilers are considered:  one in which the spoilers are 
capable of tolerating a certain number of damaging interceptions 
per collider run ("consumable" spoilers), and one in which the
spoilers are potentially damaged on every machine pulse and are
self-repairing ("renewable" spoilers).  The collimation efficiency
of the system is evaluated, considering both halo particles which
are rescattered into the beam and muon secondaries which are
passed to the interaction region.  We conclude that the new
design is a promising candidate for the NLC post-linac system.
\end{abstract}

\section{Introduction}

The experience of the Stanford Linear Collider (SLC) indicates that
collimation of the beam halo at the end of the main linacs of the
Next Linear Collider (NLC) will be a necessity.  The principal 
requirements on the NLC post-linac collimation system are as follows:
\begin{itemize}
\item The system should stop particles which would generate unacceptable
   backgrounds in the detector from entering the final focus
\item The collimation efficiency should be sufficiently high that the
   number of halo particles which are transmitted to the final focus is
   comparable to the number generated by beam-gas and thermal-photon
   scattering from the collimation region and the final focus
\item The number of muon secondaries from the collimation system which 
   reach the detector must be minimized
\item The optical and wakefield dilutions of the beam emittances due to
   the collimation system must be small
\item The system must protect the final focus and the detector from
   beams which have large energy or betatron excursions without being
   destroyed in the process.
\end{itemize}
\begin{figure}[htb]
\centering
\includegraphics*[width=60mm,angle=-90]{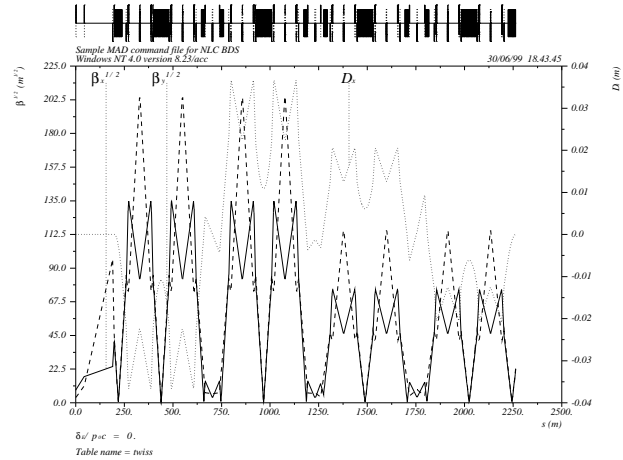}
\caption{Optical functions of 1996 NLC post-linac collimation system.}
\label{collbetas}
\end{figure}

The 1996 NLC design included a post-linac collimation system shown in
Figure \ref{collbetas} \cite{zdr}.  The system design was driven primarily
by the machine protection requirement that a single bunch train (80 kJ
at 500 GeV per beam) at nominal emittances ($\gamma\epsilon_{x,y} =
(4\times0.06)$ mm.mrad) should not be able to damage the collimators.
This required a scheme of optically-thin spoilers and thick absorbers in
each plane, large betatron functions, and strong optics, which in turn
introduced difficulties due to nonlinearities and wakefields.

The difficulties envisioned in the operation of the collimation system
led to reconsideration of the design assumptions and a new conceptual
design.

\section{Design Assumptions}

The design of the post-linac collimation system is most strongly governed
by the expected properties of large excursions which can impact the
collimators.  Previously it had been assumed that neither energy nor
betatron excursions could be trapped actively in the NLC due to its low
repetition rate (120 linac pulses per second).  A re-examination of the
SLC operational history, as well as that of other accelerators, indicated
that failures which could cause a fast (inter-pulse) betatron
oscillation of the required magnitude were either rare
or could be eliminated by design, while pulse-to-pulse energy
variations of the required magnitude cannot be ruled out for a linac.  

The expected charge of the beam halo was originally $10^{10}$ particles per
linac pulse (1\% of the beam), based on early SLC experience.  Later
SLC experience showed that the halo could be reduced substantially through
careful tuning of the injection (damping ring and compressor) systems.
In the present NLC design a collimation system downstream of the damping
ring and first bunch compressor is expected to dramatically reduce the
halo intensity at the end of the main linac.
The present estimate of the
halo is $10^7$ particles per pulse; we have chosen to design for a safety
factor of 100 over this estimate.  
This reduction eliminates the requirements
for water cooling in the spoiler elements and eases the tolerances on muon 
generation.

\section{New Collimation System Optics}

\begin{figure}[htb]
\centering
\includegraphics*[width=60mm,angle=-90]{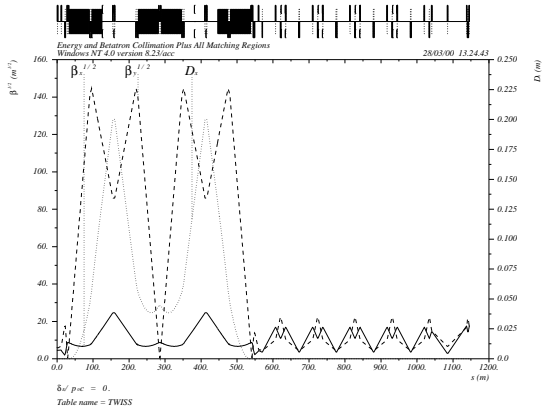}
\caption{Optical functions of proposed new NLC post-linac collimation system.}
\label{newtwiss}
\end{figure}

Figure \ref{newtwiss} shows the optical functions of the new post-linac
collimation lattice.  The energy and betatron collimators are 
separated, with the former preceding the latter.

\subsection{Energy Collimation}

The energy collimation section achieves passive protection against 
off-energy pulses through a 0.5 radiation length (R.L.) spoiler and
a 20 R.L. absorber separated by approximately 30 meters.  The first few
R.L. of the absorber are titanium, for which the RMS beam size
$\sigma_r \equiv \sqrt{\sigma_x\sigma_y}$ must be larger than 560 $\mu$m
to ensure survival \cite{zdr2}.  Beams which pass through the spoiler
will develop RMS scattering angles of 19 $\mu$radians in horizontal and
vertical; combined with the dispersive
beam size at the absorber ($\eta\sigma_{\delta} = 500 \mu{\rm m}$), the
expected size of a beam at the absorber which first passes through the
spoiler is 660 $\mu$m.  

Survival of the 0.5 R.L. spoiler is also a consideration.  At the spoiler
location in the energy collimation region, $\sigma_r = 89 \mu$m.  For the
NLC bunch train at 500 GeV per beam, the minimum beam size for survival of
a 0.5 R.L. beryllium spoiler is approximately 50 $\mu$m, thus we
have chosen beryllium as the material for the spoilers \cite{wrn}.

The collimation depth in energy should be narrower than the bandpass over
which beams are well-behaved in the final focus.  The present system is
designed to remove $\pm 1\%$ off-energy particles, which requires a half-gap
of 1.3 mm for the spoilers and 2.0 mm for the absorbers.

The jitter amplification effect of collimator wakefields must be minimized
at all points in the collimation system.  In the energy collimation 
region, the ratio $\eta_x/\beta_x$ is large and thus the collimator wakefields
primarily couple energy jitter into horizontal position jitter.  This aberration
is cancelled by placing a second spoiler-absorber pair at a location which is
$-I$ in betatron optics from the first pair but with equal dispersion functions.
The cancellation is only exact for
on-energy particles, but the expected energy jitter of 0.22\% only causes a horizontal
jitter of 0.5\% of $\sigma_x$.  A similar effect is caused by high-order
dispersion, but the effect is
approximately 1/3 as large as the residual wakefield jitter contribution.

\subsection{Betatron Collimation}

Because large betatron oscillations are not expected to develop during one
inter-pulse period, it is expected that the betatron collimators will rarely
be hit by the beam core.  The baseline design for the betatron collimation
system, which is the system pictured in Figure \ref{newtwiss},
utilizes ``consumable'' spoilers, in which the spoilers can be moved
to present a fresh surface to the beam after every incident of beam-core
interception; we assume that 1,000 such incidents can occur per year of
operation.  An alternative
design would permit damage on every pulse and require that the collimators
be self-repairing, ``renewable'' collimators.  While more techincally challenging,
the renewable collimators would permit smaller apertures to be used, which
in turn would permit smaller betatron functions.

The system in Figure \ref{newtwiss} is based on a triplet
lattice with phase advances of $\pi/2$ and $3\pi/2$ per
cell in horizontal and vertical, respectively.  Thus the system collimates
in two phases, two planes, two iterations per phase/plane.
Each high-beta region in the system contains 2 adjustable
spoilers ($x$ and $y$) and 2 fixed cylindrical absorbers.  
Multiple coulomb scattering in the spoilers gives the halo
a large angular divergence, which causes particles to hit the absorbers
in the next cell.

The required collimation aperture is set by acceptable limits on synchrotron
radiation in the final doublet. 
Based on studies of the 1996 final focus \cite{stan},
the nominal spoiler half-gaps are approximately
200 $\mu$m for 500 GeV beams.  The fixed absorbers have a round aperture
with a radius of 1 mm.  Spoilers and absorbers are 0.5 and 20.0 R.L., respectively.

The vertical 
jitter amplification factor for the betatron collimation system is 46\%,
smaller than the 66\% expected for the 1996 design.  For the expected incoming
jiter (0.375 $\sigma_y$), the collimators contribute 
0.17 $\sigma_y$ jitter in quadrature with the incoming jitter.  These estimates
are based on analytic models for collimators with a $z$ taper and a large
$x/y$ aspect ratio \cite{gennady}; however, recent experiments indicate that
the actual wakefield effect may be smaller than this \cite{collwake}.  The
horizontal jitter amplification is expected to be about half that of the
vertical.

%
%
\section{Scattering Studies}

The efficiency of primary-particle collimation and the
production of muons which are transmitted to the IP were studied using
a combination of TURTLE, EGS, and MUCUS (MUltiple CoUlomb Scattering program).

\subsection{Primary Particles}

\begin{figure}[htb]
\centering
\includegraphics*[width=60mm,angle=90]{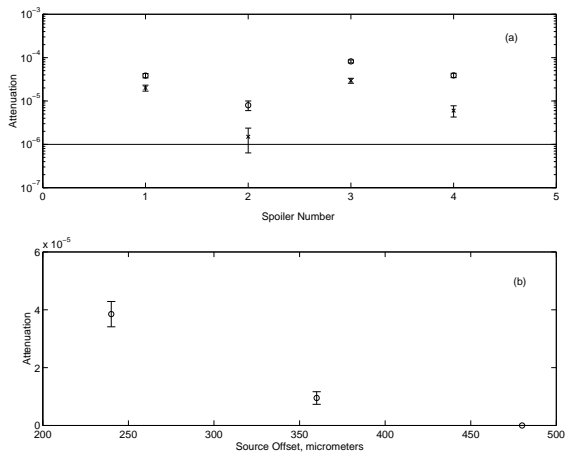}
\caption{Collimation efficiency of betatron collimation system.  (a):
   attenuation for first $y$, first $x$, second $y$, second $x$ spoiler,
   respectively, both with (circles) and without (crosses) final energy
   collimation;
   solid line shows the desired attenuation.
   (b):  Attenuation as a function of primary particle
   offset from beam axis, first $y$ spoiler.}
\label{attn}
\end{figure}

Figure \ref{attn} shows the halo attenuation based on tracking of 2 million
halo particles which originate at a point on one collimator.  Figure 
\ref{attn} (a) shows the attenuation for particles 240 $\mu$m from the
beam axis at each of the first 4 spoilers (2 vertical, 2 horizontal); the
attenuation is shown for cases in which off-energy primary particles are
collimated downstream of the collimation system (eliminating particles which
are more than 2\% off-energy), and cases without downstream energy attenuation.
The attenuation is typically between $0.6\times10^{-5}$ and $8\times10^{-5}$,
while the desired value is $0.1\times10^{-5}$.  Figure \ref{attn} (b) shows
the attenuation as a function of source offset for the first spoiler.

Note that these estimates are preliminary, and recent studies have indicated
that a substantial improvement may be achieved by optimizing the $z$ positions
of the spoilers.  Also, increasing the spoiler thickness to 1.0 R.L. would
improve attenuation by an order of magnitude, but the resulting energy
deposition in the spoilers would have to be studied. 

\subsection{Muon Secondaries}

The problem of muon secondaries from the post-linac collimators entering the
detector is more severe than it was for the 1996 NLC design, primarily because
the collimation system and final focus are shorter in the present design
(2.5 km per side compared to 5.2 km per side), which puts all sources of muons
closer to the IP.  In the present design, we include two large magnetized
toroids for muon attenuation on each side of the IP; despite this, we expect
on the order of several hundred muons per linac pulse to enter the muon
endcap of a detector similar to the ``LCD Large'' design \cite{lcdlarge},
as well as tens of muons per linac pulse in the electromagnetic calorimeter.
These studies were performed for 500 GeV beams; for 250 GeV beams, 2 orders
of magnitude improvement are expected.  The muon rate can also be reduced by
adding additional spoilers, reducing the halo intensity, or constructing a
smaller detector (such as ``LCD Small'').  Since the 500 GeV center-of-mass
(CM) results are quite acceptable, a reasonable approach to the muon situation
might be to build the system with 2 muon toroids and spaces allocated for
additional toroids, to be added in later years if required.

\section{Conclusions and Future Directions}

We have presented an optics design for the Next Linear Collider post-linac
collimation system which addresses the difficulties in the previous system
design.  The new system has weaker optics, looser tolerances, larger
bandwidth, and better wakefield properties than the original.  The new
system is somewhat poorer than desired in the areas of halo attenuation and
muon production; future work will seek to address this weakness..

The present energy collimator includes a 5 milliradian arc, which
changes the angle between the linac and the final focus.  Recent developments
in final focus design have expanded the potential energy reach of the NLC 
\cite{panta_andrey}; in order to take full advantage of this change, we plan
to redesign the collimation system to either a dogleg or a chicane, such that
the post-linac system and the linac are co-linear and the former can be
expanded by ``pushing back'' into the latter when the linac gradient and
energy are increased.

\section{Acknowledgements}

This work would not have been possible without the ideas and assistance of
G. Bowden, J. Frisch, J. Irwin, T. Markiewicz, N. Phinney and F. Zimmermann.

%

\end{document}